\documentclass[pra,aps,twocolumn,longbibliography,amsmath,amsmath,amssymb,showkeys]{revtex4-1}
\usepackage{graphicx}
\usepackage{dcolumn}
\usepackage{bm}
\usepackage{docs}
\usepackage[dvipsnames]{xcolor}
\usepackage{newtxtext,newtxmath}
\usepackage[T1]{fontenc}

\begin{document}

\title{An interacting molecular cloud scenario for production of gamma-rays and neutrinos from MAGIC J1835$-$069, and MAGIC J1837$-$073}

\author{Prabir Banik}
\email{pbanik74@yahoo.com}
\affiliation{High Energy and Cosmic Ray Research Centre, University of North Bengal, Siliguri 734013, India}

\author{Arunava Bhadra}
\email{aru\_bhadra@yahoo.com}
\affiliation{High Energy and Cosmic Ray Research Centre, University of North Bengal, Siliguri 734013, India}


\begin{abstract}
Recently the MAGIC telescope observed three TeV gamma-ray extended sources in the galactic plane in the neighborhood of radio SNR G24.7+0.6. Among them, the PWN HESS J1837$-$069 was detected earlier by the HESS observatory during its first galactic plane survey. The other two sources, MAGIC J1835$-$069 and MAGIC J1837$-$073 are detected for the first time at such high energies. Here we shall show that the observed gamma-rays from the SNR G24.7+0.6 and the HESS J1837$-$069 can be explained in terms of hadronic interactions of the PWN/SNR accelerated cosmic rays with the ambient matter. We shall further demonstrate that the observed gamma-rays from the MAGIC J1837$-$073 can be interpreted through hadronic interactions of runaway cosmic-rays from PWN HESS J1837$-$069 with the molecular cloud at the location of MAGIC J1837$-$073. No such association has been found between MAGIC J1835$-$069 and SNR G24.7+0.6 or PWN HESS J1837$-$069. We have examined the maximum energy attainable by cosmic-ray particles in the SNR G24.7+0.6/ PWN HESS J1837$-$069 and the possibility of their detection with future gamma-ray telescopes. The study of TeV neutrino emissions from the stated sources suggests that the HESS J1837$-$069 should be detected by IceCube Gen-2 neutrino telescope in a few years of observation.
\end{abstract}

\maketitle

\section{Introduction}
While studying a star-forming region in the inner part of the galactic plane centred around the supernova remnant (SNR) G24.7+0.6 (3FHL J$1834.1-0706e$/ FGES J$1834.1-0706$), which a middle-aged composite SNR, the Major Atmospheric Gamma Imaging Cherenkov (MAGIC) Telescopes recently discovered a new extended gamma-ray source MAGIC J1835$-$069 at energies above $150$ GeV to till 5 TeV in the neighborhood of the supernova remnant (SNR) G24.7+0.6 \cite{MAGIC18}. Besides, the MAGIC telescope also detected high energy gamma rays from two other sources in the region namely HESS J1837$-$069, and MAGIC J1837$-$073. In fact, the HESS J1837$-$069 is the brightest object in the region which was discovered by the High Energy Stereoscopic System (HESS) observatory during its first galactic plane scan \cite{Aharonian06}. The MAGIC J1837$-$073 is adjacent to HESS J1837$-$069 and coincides with the HESS hotspot in the south of HESS J1837$-$069.

The MAGIC observations suggest that the center position of the extended object MAGIC J1835$-$069 is located $0.34^{\circ}$ away from the center of the SNR G24.7+0.6. The spectral studies of the stated two sources suggest that the statistical significance for a common origin of the MAGIC J1835$-$069 and the SNR G24.7+0.6 is less than $1.5\sigma$. The gamma-ray spectral measurements done by Fermi-LAT and MAGIC observatories suggest that the two sources are likely to be correlated \cite{MAGIC18}. The observation of $^{13}$CO J $= 1-0$ line at 110 GHz favors for molecular cloud origin of observed gamma-rays from MAGIC J1835$-$069 \cite{Jackson06,Petriella12}.  

The gamma-ray extended source HESS J1837$-$069 was initially classified as an unidentified object \cite{Aharonian06}. It coincides with the x-ray source AX J1838.0-0655. A young and energetic 70.5 ms rotation powered pulsar PSR J1838-0655 in the region was discovered by the Rossi X-Ray Timing Explorer (RXTE) \cite{Gotthelf08}. The high resolution of Chandra observation revealed that AX J1838.0-0655 contains a bright point source surrounded by a centrally peaked nebula \cite{Gotthelf08}. Following these discoveries, the source HESS J1837$-$069 is interpreted as a pulsar wind nebula \cite{Fujita14,Sun20}. The size of the AX J1838.0-0655 in x-rays is smaller than the HESS J1837$-$069 observed in gamma rays which are generally interpreted as the diffusion of high energy particles in the surrounding medium. The origin of GeV-TeV gamma rays from MAGIC J1837$-$073 is so far not discussed in the literature.

Under the circumstances, the main objectives of the present work are as follows: First, we would like to examine whether the spectral behavior of observed gamma-rays from the HESS J1837$-$069 and SNR G24.7+0.6 can be explained by hadronic interactions of the PWN/ SNR accelerated cosmic rays with the ambient (proton) matter.  Secondly, with the same cosmic ray production spectrum which is required to explain the observed gamma-rays directly from HESS J1837$-$069 and SNR G24.7+0.6, we would like to explore whether one can consistently explain the observed EM SED from the sources MAGIC J1837$-$073 and MAGIC J1835$-$069 with a scenario of cosmic-ray-illuminated cloud interacting with the PWN HESS J1837$-$069/SNR G24.7+0.6. We would like to determine the corresponding neutrino fluxes from all the stated sources and subsequently study their detection likelihood by the present/upcoming neutrino observatories.

The plan of the paper is the following: In the next section, we shall describe the methodology for evaluating the gamma-ray and neutrino fluxes from the PWN/SNR and the molecular cloud interacting with the PWN/SNR. The gas distributions and subsequently masses of the regions under the study assuming MAGIC J1835$-$069, and MAGIC J1837$-$073 located in the vicinity of PWN HESS J1837$-$069/SNR G24.7+0.6 are estimated in the same section. In sect. 3, the numerical estimated fluxes of hadronically produced gamma-rays and neutrinos from the PWN/SNR and surrounding molecular clouds over the GeV to TeV energy range along with the observations. The findings of the present work will be discussed in Sect. 4 and we shall conclude finally in the same section.

\section{The gamma rays and neutrinos production methodology}
\label{sec:1}
The production spectrum of cosmic rays at PWN/SNR can be represented by a power law and is given by
\begin{eqnarray}
 \frac{dn_p}{dE_p} = K E_p^{-\alpha} .
\label{cr_prod}
\end{eqnarray}
where $\alpha$ is the spectral index, $E_p$ is the energy of the cosmic rays. The normalization constant $K$ is related with the supernova explosion energy $E_{sn}$  via the relation \cite{Banik17}
\begin{eqnarray}
\xi E_{sn} =  \displaystyle\int_{E_{p,min}}^{E_{p,max}}E_p \frac{dn_p}{dE_p} dE_p
\label{Lcr}
\end{eqnarray}
where $\xi$ denotes the efficiency of supernova explosion energy into cosmic rays, $E_{p,min}$ and $E_{p,max}$ represents the minimum and maximum energies of accelerated cosmic ray protons.

The high energy gamma-ray flux from the PWN/SNR can be explained by the interaction of the shock accelerated cosmic rays with the ambient matter (protons) of density $n_{H}$. Such interaction will produce gamma-rays and neutrinos after subsequent decay of neutral and charged pions respectively along with the other secondary particles. Here we follow Banik \& Bhadra (2017) \cite{Banik17} to estimate the differential flux of high energy gamma rays reaching the Earth from the SNR G24.7+0.6. The corresponding flux of high energy neutrinos reaching the Earth from the direction of the SNR can be written as
\begin{eqnarray}
\frac{d\Phi_{\nu}}{dE_{\nu}}(E_{\nu}) = \frac{1}{4\pi d^{2}}Q_{\nu}^{pp}(E_{\nu})
\end{eqnarray}
where $Q_{\nu}^{pp}(E_{\nu})$ represents the resulting neutrino emissivity which can be obtained by following \cite{Kelner06,Banik18} and $d$ denotes the distance between the PWN/SNR and the Earth.

After escaped away from the PWN/SNR, the accelerated cosmic ray protons diffuse in the interstellar medium with a diffusion coefficient given by \cite{Giuliani10} $D = D_0(\frac{E_p}{10 GeV})^{\delta}$ where $D_0$ represents the diffusion coefficient of accelerated protons at 10 GeV energy ($\sim 10^{28}$ cm$^2$s$^{-1}$ in our Galactic medium) and $\delta$ indicates a constant having value between $0.3$ to $0.7$ \cite{Berezinsky90,Strong07}. However, a slow diffusion has been inferred in the dense gaseous medium of molecular clouds \cite{Aharonian96,Ormes88}. The measurement of Boron to Carbon Flux Ratio in Cosmic Rays over the rigidity extent of 1.9 GV to 2.6 TV by the Alpha Magnetic Spectrometer (AMS-02) on the International Space Station found that $\delta$ is around $0.33$ \cite{Aguilar16}. Becker et al. (2016) \cite{Becker16} have recently demonstrated that the cosmic ray budget can be likened well for a diffusion coefficient that is close to $D \propto E^{0.3}$ by studying 21 SNRs, those are well-studied from radio wavelengths up to gamma-ray energies. On the other hand, if cosmic rays originated at SNRs of the galaxy, a more powerful diffusion with $\delta = 0.5$ cannot generate the observed cosmic ray energy spectrum especially at the high-energy (TeV) component of the spectrum \cite{Becker16}. The probability density of finding a cosmic ray particle at a given radius $r$ from a PWN/SNR for an impulsive injection spectrum, after including the energy loss due to hadronic $pp-$interaction of cosmic rays in PWN/SNR, can be written as \cite{Aharonian96,Bhadra02,Torres08} 
\begin{equation}
\scriptsize
 P(E_p,r,t) = \frac{1}{\pi^{3/2} r_{dif}^{3} } \exp(-(\alpha- 1)t/\tau_{pp} - (r/r_{dif})^2)
\label{prob_den}
\end{equation}
where $t$ denotes the age of the supernova explosion/PWN, $\tau_{pp} = n_H ck\sigma_{pp}$ stands for represents the cosmic ray energy loss time-scale due to hadromic interaction of cosmic rays with ambient matter of the PWN/SNR, $k \sim 0.45$ denotes the inelasticity of the interaction, and $\sigma_{pp}$ \cite{Kelner06} being the interaction cross-section \cite{Torres08}. Here, $r_{dif} = 2 \sqrt{ D t [\exp(t\delta/\tau_{pp})-1]/[t\delta/\tau_{pp}]}$ is the radius of the sphere up to which the cosmic ray particles have time to travel after their injection \cite{Torres08}. Therefore, the intensity of cosmic rays at a distance $r$ from the source can be written as \cite{Banik18}
\begin{eqnarray}
 \frac{dn_p'}{dE_p}(r,t) = P(E_p,r,t)\frac{dn_p}{dE_p}
\label{CR_den_cl}
\end{eqnarray}
where it is assumed that cosmic rays are diffusing from a point source.

To explain the MAGIC collaboration observed high energy gamma-rays from the source MAGIC J1835$-$069 and MAGIC J1837$-$073, we consider a dense molecular cloud at the center of this source. The cloud is continuously illuminated by runaway relativistic cosmic ray particles accelerated at the nearby SNR G24.7+0.6 or the PWN HESS J1837$-$069. We follow Banik \& Bhadra (2017) \cite{Banik17} and Banik \& Bhadra (2018) \cite{Banik18} to find out the differential flux of high energy gamma rays and neutrinos reaching the Earth from the source MAGIC J1835$-$069 and MAGIC J1837$-$073 after their production in the interaction of runaway relativistic cosmic ray particles with molecular clouds.

Since our objective is to examine whether the observed gamma rays from MAGIC J1835$-$069 and MAGIC J1837$-$073 are correlated with the two known high energy gamma rays SNR/PWN of nearby regions (SNR G24.7+0.6 and HESS J1837$-$069), we exclude the leptonic origin scenario of gamma rays from SNR G24.7+0.6 and HESS J1837$-$069. The decay of charged pions produced in pp interactions at the molecular cloud region will finally lead to electrons and positrons along with the neutrinos. The so produced electrons will give synchrotron photons while traveling through the magnetic field of the molecular cloud. However, the density of soft photons in the cloud region (synchrotron photons) will be low because of the weak magnetic field of the clouds. As a result, the inverse Compton process is not significant in the present scenario.   

\subsection{Gas density map around SNR G24.7+0.6/PWN HESS J1837$-$069}
The gas distribution map of the regions, particularly in the locations of SNR G24.7+0.6, PWN HESS J1837$-$069, MAGIC J1835$-$069, and MAGIC J1837$-$073 is studied below 
considering that either MAGIC J1835$-$069, and MAGIC J1837$-$073 are at the vicinity of PWN HESS J1837$-$069 or of SNR G24.7+0.6. 

The interstellar molecular clouds are a complex mixture of gas at certain velocities. The major constituent of molecular clouds is hydrogen which may remain in three different phases, the neutral atomic hydrogen (H{\small I}), the ionized hydrogen (H{\small II}), and the molecular hydrogen (H$_2$). The 
cold H$_2$ is not directly observable in emission. Carbon monoxide (CO) is widely used as tracer for H$_2$, the intensity of rotational transition of CO relates with the abundance of H$_2$. The column density of atomic hydrogen can be estimated straightway from the 21 cm radio frequency transition between the two hyperfine levels of the ground state of H{\small I}. The regions of H{\small II} is characterized by thermal free-free radiation and recombination lines. 

\begin{figure}
    \includegraphics[width = 0.5\textwidth,height = 0.4\textwidth]{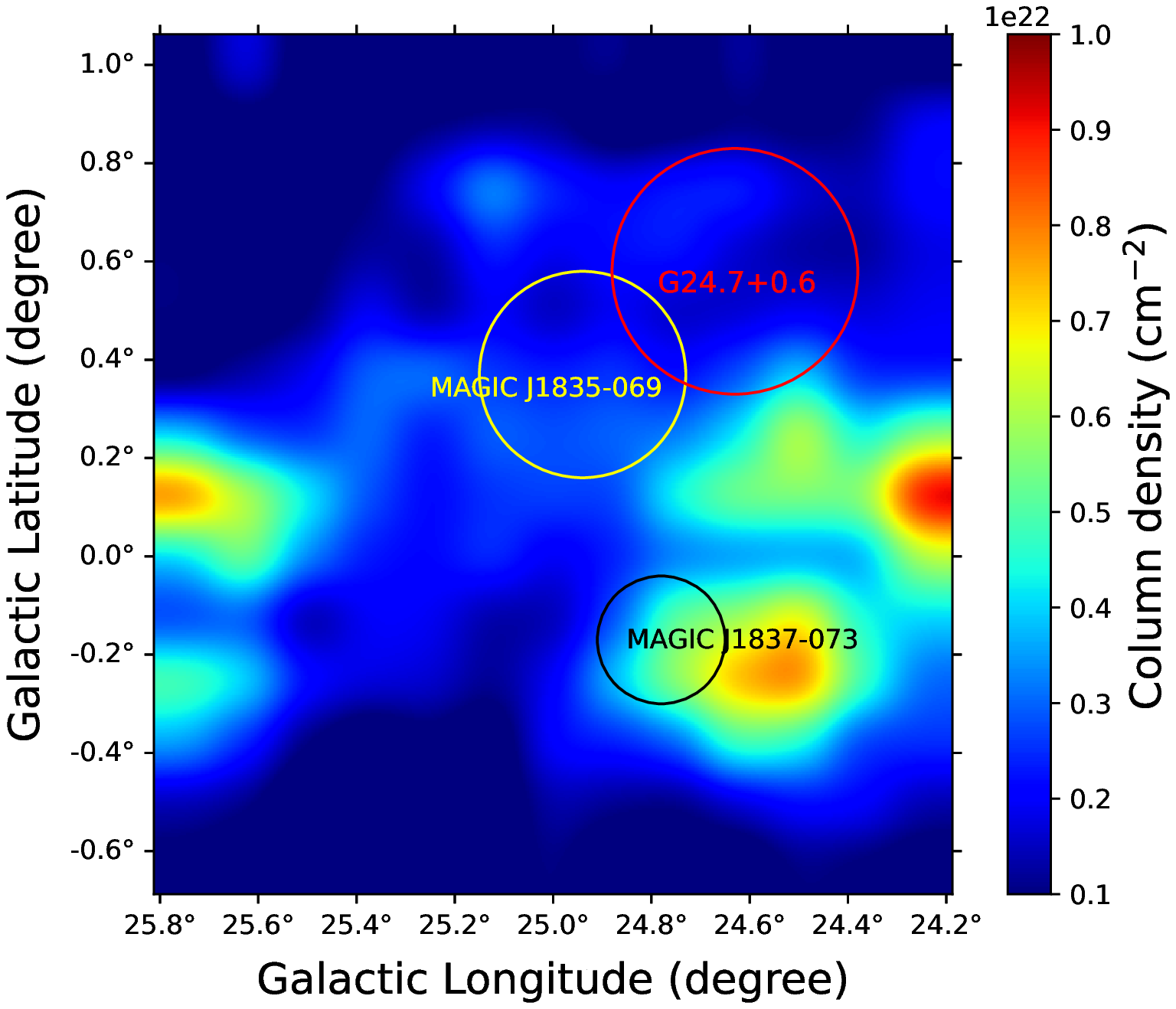}
    \includegraphics[width = 0.5\textwidth,height = 0.38\textwidth]{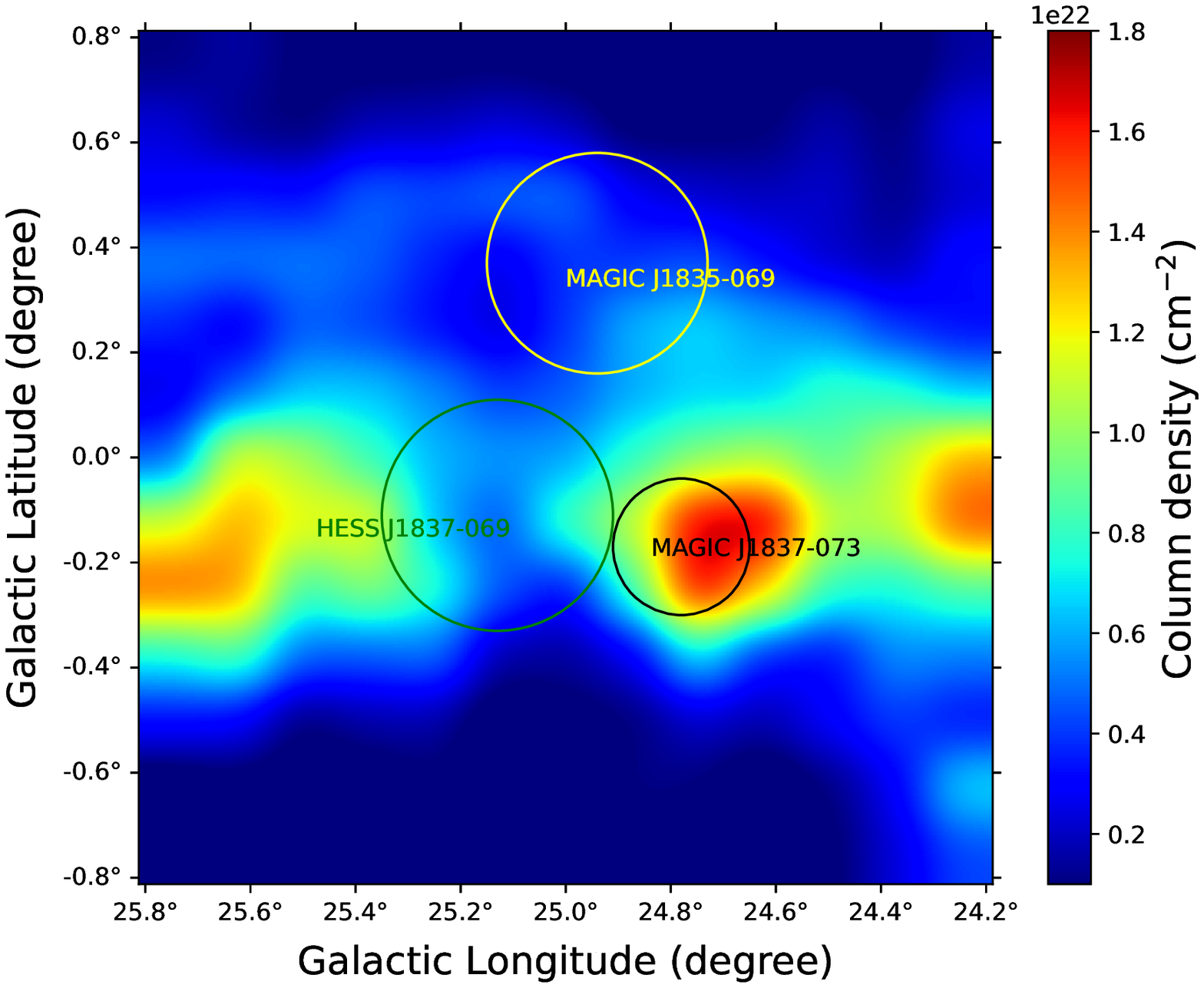}
 \caption{$H_2$ column density obtained from the carbon monoxide (CO) composite survey. Top: $H_2$ column density of gas content integrated over the velocity interval from 21 to 41 km s$^{-1}$ which corresponds to 3.5 kpc. Bottom: $H_2$ column density of gas content integrated over the velocity interval from 78 to 98 km s$^{-1}$ which corresponds to 6.6 kpc.}
\label{Fig:1}
\end{figure}

To estimate the column density for H$_2$, we take data for $^{12}$CO $J = 1-0$ emission from the carbon monoxide (CO) composite survey \cite{Dame01}. We evaluate the column density of molecular hydrogen by assuming a linear relationship $N_{H_2} = X_{CO} \times W_{CO}$, where $W_{CO}$ represent the velocity-integrated brightness temperature of CO 2.6-mm line \cite{Katsuta17,Bolatto2013}. Here, the conversion factor, $X_{CO}$ is assumed to be $3 \times 10^{20}$ cm$^{-2}$ K$^{-1}$ km$^{-1}$ s which is consistent with the suggested conversion factor of $X_{CO} \approx (2-4) \times 10^{20}$ cm$^{-2}$ K$^{-1}$ km$^{-1}$ s from different observations for the molecular clouds very near to galactic plane \cite{Dame01,Narayanan11}. We use the circular rotation model for the milky-way Galaxy as given in Ranasinghe \& Leahy (2018) \cite{Ranasinghe18} and the rotational curve data for gas velocities at different Galactocentric distance \cite{Sofue11} to correlate measured radial velocity of gases with kinematic distance. We obtained a radial velocity of the molecular hydrogen gas about 88 km s$^{-1}$ and 31.2 km s$^{-1}$ which corresponds to 6.6 kpc and 3.5 kpc distance from us respectively. Therefore we choose a radial velocity range of $v = 78-98$ km s$^{-1}$ (or $v = 21-41$ km s$^{-1}$) to be integrated to determine the column density of molecular hydrogen at a distance of about 6.6 Kpc (or 3.5 Kpc) and obtained column density map of gases are shown in the Fig.~\ref{Fig:1}.


We obtain the column density of H{\small II} from the Planck observed free-free emission at 30 GeV frequency \cite{Planck16}. First, we use the conversion factor at 30 GHz as given in table 3 of Planck Collaboration X 2016 \cite{Planck16} to convert the measured brightness temperature into free$-$free intensity ($I_{\nu}$). Then we follow the equation (5) in Sodroski et al. (1997) \cite{Sodroski97} to transform the free$-$free intensity into column density. To calculate the H II column density, we adopted an electron temperature of $T_e = 8000$ K and the effective density of electrons of $n_e = 10$ cm$^{-3}$ \cite{Sodroski97}.


The column density of H{\small I} in the aforementioned regions is estimated under the consideration of the optically thin limit \cite{Katsuta17,Dickey90} from the data cube of Galactic H{\small I} $4\pi$ survey (HI4PI) \cite{HI4PI}. To evaluate gas column density distributions of H{\small I}, we integrated the gases over the radial velocities $v = 78-98$ km s$^{-1}$ for 6.6 kpc and $v = 21-41$ km s$^{-1}$ for 3.5 kpc.

The total mass within a molecular cloud/source has been estimated from the expression \cite{Sun20}
$$M_H = m_H N_{H_i} A_{ang} d^2$$
where $m_H$ represents the mass of the hydrogen atom, $N_{H_i}$ is the column density of hydrogen in ith phase ($i \equiv  2, I, II$ ), $A_{ang}$ denotes to the angular area, and $d$ indicates the distance of the molecular cloud/source. 


The total mass of hydrogen gas content of the sources SNR G24.7+0.6 (at 3.5 kpc), HESS J1837$-$069 (at 6.6 kpc) and two possible associated sources, i.e  MAGIC J1835$-$069 and MAGIC J1837$-$073 either at 3.5 kpc or 6.6 kpc are shown in Table~\ref{table11} and Table~\ref{table22} respectively. 


\begin{table}
    \caption{The mass of the gas content in the sources at 3.5 kpc.}
    \label{table11}
    \begin{tabular}{llll}
 \hline\noalign{\smallskip}
 & \multicolumn{3}{c}{ Mass of the gas content in  M$_{\odot}$ for} \\ \cline{2-4} 
\noalign{\smallskip}
          Source         &  H$_2$ & H {\small I}&  H {\small II}    \\ 
\noalign{\smallskip}\hline\noalign{\smallskip}
       SNR G24.7+0.6    & $2.5\times 10^{4}$    &  $1.5\times 10^{4}$  & $1.3\times 10^{4}$  \\
      MAGIC J1835$-$069   & $2.3\times 10^{4}$    &  $1.1\times 10^{4}$ &  $1.2\times 10^{4}$  \\ 
      MAGIC J1837$-$073   & $1.2\times 10^4$    & $4.1\times 10^3$ & $5.5\times 10^{3}$\\
\noalign{\smallskip}\hline    
    \end{tabular}
\end{table}

\begin{table}
    \caption{The mass of the gas content in the sources at 6.6 kpc.}
    \label{table22}
    \begin{tabular}{llll}
 \hline\noalign{\smallskip}
 & \multicolumn{3}{c}{ Mass of the gas content in  M$_{\odot}$ for} \\ \cline{2-4} 
\noalign{\smallskip}
          Source         &  H$_2$ & H {\small I}&  H {\small II}    \\ 
\noalign{\smallskip}\hline\noalign{\smallskip}
      HESS J1837$-$069    & $1.9\times 10^{5}$    &  $3.8\times 10^{4}$  & $5.6\times 10^{4}$  \\
      MAGIC J1835$-$069   & $1.1\times 10^{5}$    &  $3.7\times 10^{4}$ &  $4.3\times 10^{4}$  \\ 
      MAGIC J1837$-$073   & $2\times 10^5$    & $1.5\times 10^4$ & $1.9\times 10^{4}$\\
\noalign{\smallskip}\hline    
    \end{tabular}
\end{table}

\section{Results and Discussion}

Before the report by the MAGIC collaboration, Katsuta et al. (2017) \cite{Katsuta17} carried out a study of the complex region around the SNR G24.7+0.6 with the Fermi-LAT telescope and found GeV gamma-ray emission from two elliptical extended regions G25A and G25B, each composed of three components. Note that, the emission from the G25A1 component is coincident with the new source MAGIC J1835$-$069 detected by MAGIC \cite{MAGIC18}. In addition, through X-ray observations of the region with XMM$-$Newton \cite{Katsuta17} found a young massive OB association/ cluster, G25.18+0.26 which is proposed to be associated with the extended GeV emissions (G25A and G25B). According to the MAGIC collaboration the emission detected at VHE with MAGIC is unlikely to be correlated with the OB association/ cluster G25.18+0.26 detected in X-rays as MAGIC only detects emission from the G25A1 component \cite{MAGIC18}. As mentioned above we have estimated high energy gamma rays to explain the observed gamma-ray spectrum from the sources, SNR G24.7+0.6 and HESS J1837$-$069 in the hadronic interaction of cosmic rays with ambient matter. We have also investigated below the observed gamma-ray spectrum of the sources, MAGIC J1835$-$069 and MAGIC J1837$-$073 in terms of the hadronic interaction of cosmic rays with giant molecular clouds at the centre of these sources which are illuminated by the accelerated cosmic rays either from SNR G24.7+0.6 or HESS J1837$-$069. The estimation of gamma ray flux along with observed data and consequently corresponding predicted high energy neutrino fluxes from the said sources are discussed bellow.

\begin{figure}
  \begin{center}
  \includegraphics[width = 0.5\textwidth,height = 0.45\textwidth,angle=0]{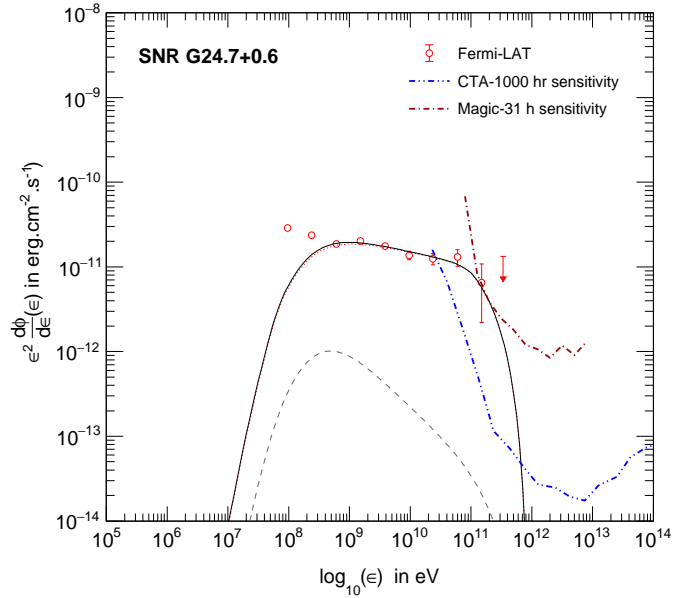}
\end{center}
  \caption{The estimated differential energy spectrum of gamma-rays reaching the Earth from the SNR G24.7+0.6. The red dotted line represents the gamma-ray flux produced due to interaction of SNR accelerated cosmic rays with the ambient matter of the SNR. The gray dotted line denotes the background gamma ray flux from the SNR due to interaction the cosmic rays galactic background with SNR material. The black continuous line represents the estimated overall differential EM SED from the SNR G24.7+0.6. The blue dash-double-dotted and brown dash-single-dotted lines represent the detection sensitivity of the CTA detector for 1000 hours and the MAGIC detector for 31 hour, respectively.}
\label{Fig:2}
\end{figure}

\begin{figure*}
  \begin{center}
  \includegraphics[width = 1.0\textwidth,height = 0.45\textwidth,angle=0]{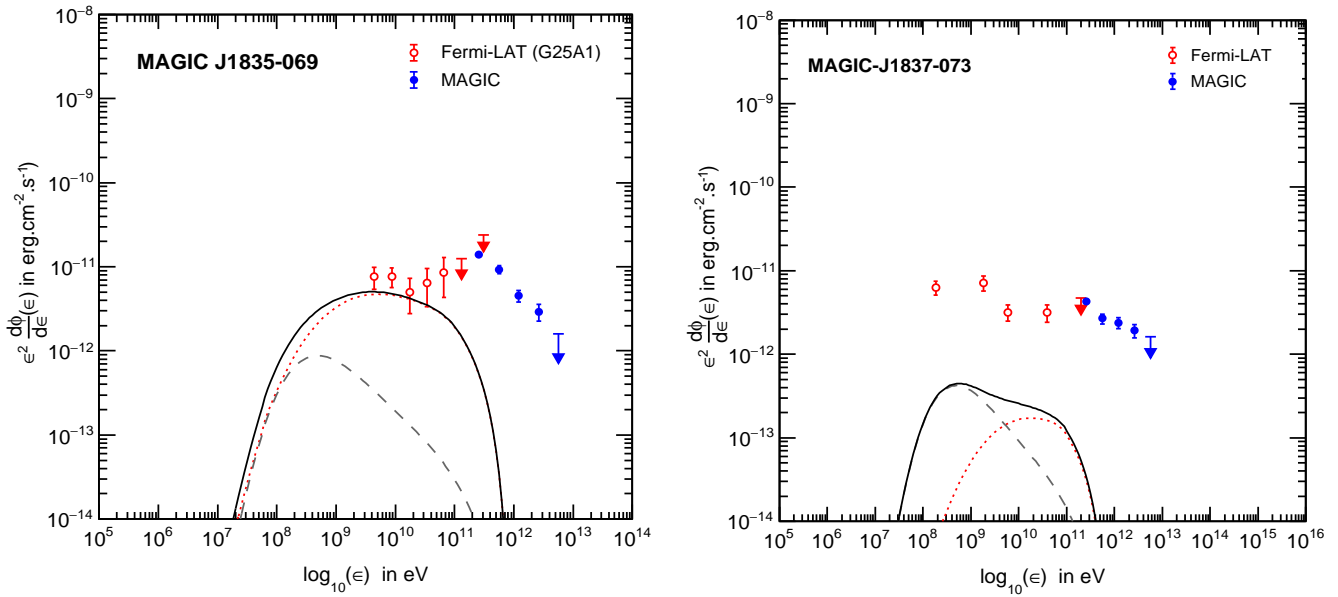}
\end{center}
  \caption{Left: The estimated gamma-ray flux reaching the Earth from the MAGIC J1835$-$069. The red dotted line represents the corresponding gamma-ray flux produced due to interaction of SNR accelerated cosmic rays with the molecular cloud. The gray dotted line denotes the corresponding background gamma-ray flux from MAGIC J1835$-$069 produced due to interaction the Cosmic rays galactic background with molecular cloud. The black continuous line represents the estimated overall differential EM SED from the MAGIC J1835$-$069. Right: The estimated gamma-ray flux reaching the Earth from the MAGIC J1837$-$073. The red dotted line represents the corresponding gamma-ray flux produced due to interaction of SNR accelerated cosmic rays with the molecular cloud. The gray dotted line denotes the corresponding background gamma-ray flux from MAGIC J1835$-$073 produced due to interaction the Cosmic rays galactic background with molecular cloud. The black continuous line represents the estimated overall differential EM SED from the MAGIC J1835$-$073.}
\label{Fig:3}
\end{figure*}

\subsection{SNR G24.7+0.6}
The SNR G24.7+0.6 is a radio and gamma-ray supernova remnant advancing in a dense medium \cite{MAGIC18,Reich84,Seward90,Leahy89,Ackermann17,Ajello17}. 
A recent study based on analysis of the 1420 MHz radio continuum data, H I-line absorption spectra, and H I channel maps which is obtained from the VLA (Very Large Array) Galactic Plane Survey (VGPS) \cite{Stil06}, suggests a new distance about $D \sim 3.5 \pm 0.2$ kpc of the SNR from us \cite{Ranasinghe18}. Using the new distance ($d$) measurement of the SNR \cite{Ranasinghe18}, we have estimated the age of the SNR to be $t = 2.09\times 10^4$ year from $Sd^2-t$ relation as given in \cite{Weiler80} where $S\;(1\; GHz) = 20\; Jy$ is the observed flux density from the SNR at 1 GHz frequency \cite{Reich84}. Using this distance of the SNR, the size of the SNR has been estimated to be $(30.5 \pm 1.7) \times (15.3 \pm 0.9)$ pc \cite{Ranasinghe18} and therefore mean radius of the SNR will be $\sim 11.45$ pc. The mean radius in the 1420 MHz radio continuum image can be considered as the shock wave radius $R$. Here we apply a basic Sedov model \cite{Cox72} and the shock wave radius in the Sedov stage can be written as
$$R = 12.9 \left(\frac{E_{sn}}{n_H}\right)^{1/5}t_4^{2/5}\; pc$$
where $n_H$ is the ambient medium (proton) density in cm$^{-3}$ units, $E_{sn}$ is the supernova explosion energy in units of $0.75 \times 10^{51}$ erg and $t_4 = t/10^4$ is the age of the SNR in units of $10^4$ year. For an age of the SNR about $t \sim 2.09\times 10^4$ year and shock wave radius $R = 11.45$ pc, we can adopt a typical supernova explosion energy of $E_{sn} = 1.8 \times 10^{51}$ erg and an ambient matter density of the SNR of $n_H = 19$ cm$^{-3}$ in order to explain observed gamma-rays from the SNR G24.7+0.6.

Here we have explored whether the observed high energy gamma-ray spectrum from the SNR G24.7+0.6 can be interpreted through hadronic interactions of cosmic rays (mainly protons) accelerated by the SNR with the ambient (proton) matter. The primary cosmic-ray production spectrum has been obtained assuming a typical supernova explosion energy of $E_{sn} = 1.8\times 10^{51}$ erg with a spectral index of $\alpha = -2.2$ and we have considered an ambient matter density of the SNR of $n_H = 19$ cm$^{-3}$ as mentioned above in order to match the observed gamma-ray spectrum. We have found that a $\sim 10\%$ efficiency of conversion of supernova explosion energy to cosmic ray (proton) energy can explain well the observed experimental data. The estimated differential gamma-ray flux reaching the Earth from the SNR is displayed in the Fig.~\ref{Fig:2} along with the observations. The MAGIC detector sensitivity for 31 hours of observation with the medium zenith angle ($30^{\circ}-45^{\circ}$) has been obtained from the MAGIC detector sensitivity for 50 hours as given in \cite{Aleksi16} by assuming $1/$observational-time dependency of the sensitivity for short observation times \cite{Aleksi16}. The non-detection of gamma-rays from the SNR G24.7+0.6 by MAGIC observatory restricts the maximum energy $E_{p, max}$ attainable by a cosmic ray proton in the SNR shock to $1$ TeV. Such a choice of maximum energy of cosmic rays leads our estimation of gamma-ray flux compatible with the Fermi-LAT observation and non-detection by the MAGIC detector (considering sensitivity of 31 h for detection of a point source with $5\sigma$ significance) as shown in the figure. Future large-area telescopes like CTA \cite{Ong17} and LHAASO \cite{Liu17}, which have unprecedented sensitivity up to 100 TeV energies, should be able to identify the maximum attainable energy up to which cosmic rays can accelerate in the supernova remnant. We have also included the effect of background gamma-ray flux produced due to the interaction of diffuse cosmic rays in the galaxy with the matter spread over by the progenitor star after a supernova explosion. Here we use our estimated mass of gas content of the SNR of $\sim 5.3\times 10^{4}$ M$_{\odot}$ to estimate the background gamma-ray flux.
The Model fitting parameters for observed gamma-rays from SNR G24.7+0.6 are shown in Table~\ref{table1}.

Because of the low value of maximum energy attainable by cosmic rays in the source, the SNR G24.7+0.6 should not generate neutrinos above 1 TeV energies. 

\subsubsection{MAGIC J1835$-$069}
The gamma-ray source MAGIC J1835$-$069 has been detected with a peak significance of $11\sigma$ and resolved from SNR G24.7+0.6 at a level of $13\sigma$ for the first time \cite{MAGIC18}. Its gamma-ray spectrum is well-represented by a power-law function with a spectral photon index of $2.74 \pm 0.08$ \cite{MAGIC18}. The source positions between two extended sources detected above 10 GeV by Fermi-LAT, one of which is FGES J1836.5$-$0652 and the other is FGES J1834.1$-$0706, which are associated with HESS J1837$-$069 and SNR G24.7+0.6 respectively. The centre of MAGIC J1835$-$069 is located at $\theta = 0.34^{\circ}$ away with respect to the center of the radio SNR G24.7+0.6. MAGIC collaboration suggested that the detected gamma-ray emission can be explained as the result of proton$-$proton interaction between the cosmic rays accelerated by SNRs and the CO-rich surrounding \cite{MAGIC18}.

To explain the observed high gamma-ray spectrum from the unknown source MAGIC J1835$-$069, we have considered MAGIC J1835$-$069 as a giant molecular cloud that is illuminated by the runaway cosmic rays accelerated in the SNR G24.7+0.6. Here we have taken the distance of MAGIC J1835$-$069 from the SNR G24.7+0.6 as $r = 30$ pc following the MAGIC observation. Here we take the mass of the molecular cloud MAGIC J1835$-$069 $4.6\times 10^{4}$ M$_{\odot}$ which is estimated considering the source distance at $3.5$ kpc to explain gamma-ray observations.

The high energy cosmic rays accelerated in a DSA mechanism at the shock front of the SNR G24.7+0.6 may escape and finally interact hadronically with the matter (dominantly protons) of molecular cloud MAGIC J1835$-$069 after a diffusive propagation over a distance of $r$. Consequently, such hadronic interaction will produce high energy gamma-rays and neutrinos from the molecular cloud after subsequent decays of neutral and charged pions. The estimated differential gamma-ray flux reaching the Earth from MAGIC J1835$-$069 is shown in the left-hand panel of Fig.~\ref{Fig:3} along with the observations. Here we have adopted a diffusion constant of $D_0 = 10^{27}$ cm$^2$ s$^{-1}$ with $\delta = 0.33$ for cosmic rays. Our findings suggest that the observed GeV-TeV gamma-rays from the location of MAGIC J1835$-$069 can not be interpreted by the presented model for any reasonable fitting parameters. The present analysis thus does not support the hypothesis of possible association of the unknown source MAGIC J1835$-$069 with the SNR G24.7+0.6. The possible correlation of MAGIC J1835$-$069 with HESS J1837$-$069 is examined below.

\subsubsection{MAGIC J1837$-$073}

The detailed production model for GeV-TeV gamma-rays from MAGIC J1837-073 (3FGL J1837.6-0717 \cite{Acero15}) is not available in the literature yet. An analysis similar to the MAGIC J1835$-$069 as reported above suggests that this unknown TeV source is not associated with the SNR G24.7+0.6. The estimated differential gamma-ray flux reaching the Earth from MAGIC J1837$-$073 is shown in the right-hand panel of Fig.~\ref{Fig:3} along with the observations where we adopt the distance of MAGIC J1835$-$073 from the SNR G24.7+0.6 to be $r = 65$ pc and a diffusion constant of $D_0 = 3\times 10^{27}$ cm$^2$ s$^{-1}$ with $\delta = 0.33$ for cosmic rays. Here we take the mass of the molecular cloud as $\sim 2.2\times 10^{4}$ M$_{\odot}$ which is estimated as stated above considering cloud distance of $3.5$ kpc.

\begin{table}
    \caption{Model fitting parameters for observed gamma-rays and predicted neutrino flux from SNR G24.7+0.6 and as well as HESS J1837$-$069.}
    \label{table1}
    \begin{tabular}{llll}
 \hline\noalign{\smallskip}
      Parameters &  SNR G24.7+0.6 & HESS J1837$-$069      \\ 
\noalign{\smallskip}\hline\noalign{\smallskip}
       $d$  (in kpc)       &   $3.5$        & $6.6 $    \\
      $E_{sn}$ (in erg)    & $1.8\times 10^{51}$   &  $3\times 10^{51}$\\
      $E_{p,max}'$ (in TeV)          & 1   &  70 \\ 
      $\xi$                &  10\%   & 10\% \\
      $\alpha$              &   $- 2.2$  &   $-1.9$   \\ 
      $n_H$  (in cm$^{-3}$) & $19$  &   $8.7$\\ 
      $t$   (in yr)     & $2.09\times 10^4$ &  $2.3\times 10^4$ \\ 
\noalign{\smallskip}\hline    
    \end{tabular}
\end{table}

\begin{figure*}
  \begin{center}
  \includegraphics[width = 1.0\textwidth,height = 0.45\textwidth,angle=0]{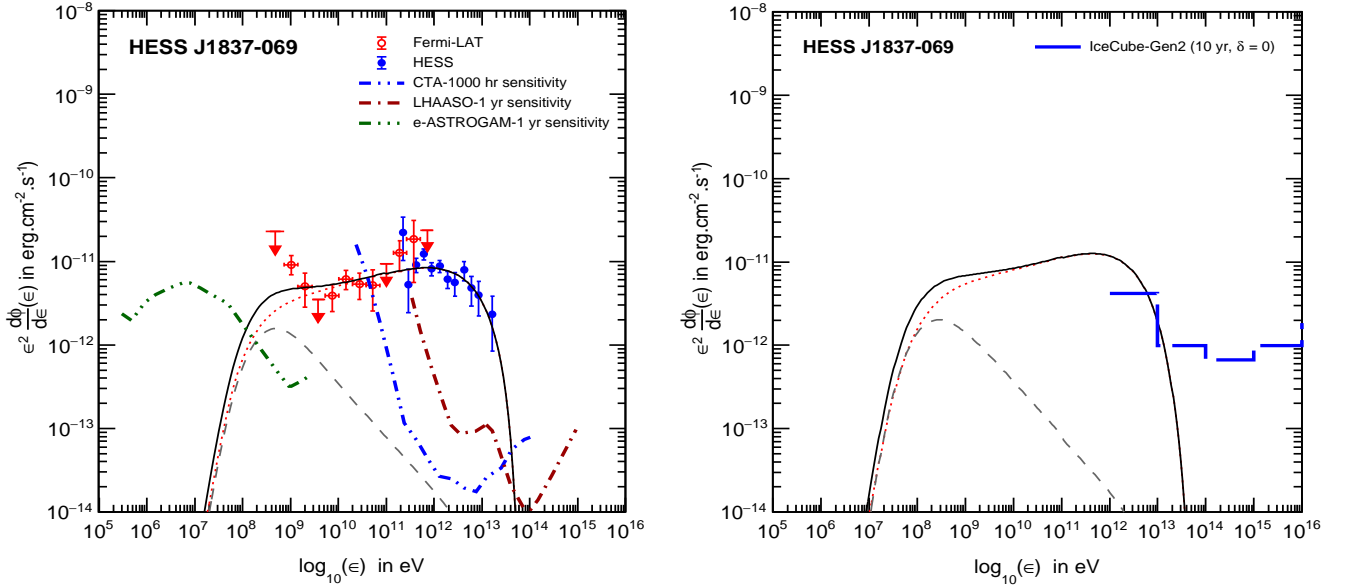}
\end{center}
  \caption{Left: The estimated differential energy spectrum of gamma-rays reaching the Earth from the PWN HESS J1837$-$069. The red dotted line represents the gamma-ray flux produced due to the interaction of SNR accelerated cosmic rays with the ambient matter of the PWN. The gray dotted line denotes the background gamma-ray flux from the PWN due to interaction of the Cosmic rays galactic background with PWN material. The black continuous line represents the estimated overall differential EM SED from the PWN. The blue dash-double-dotted, brown dash-single-dotted, and green dash-triple-dotted lines represent the detection sensitivity of the CTA detector for 1000 hours, the LHAASO detector for one year, and e$-$ASTROGAM for one year, respectively.  Right: The estimated neutrino flux reaching the Earth from the PWN HESS J1837$-$069. The red dotted line represents the corresponding all flavor neutrino flux produced due to the interaction of SNR accelerated cosmic rays with the molecular cloud. The gray dotted line denotes the corresponding background of all flavor neutrino flux from PWN HESS J1837$-$069 produced due to interaction of the cosmic rays galactic background with a molecular cloud. The black continuous line represents the estimated overall differential of all flavor neutrino flux from the PWN. Here blue long-dashed line indicates the sensitivity of IceCube-Gen2 to detect the neutrino flux from a point source at the celestial equator with an average significance of $5\sigma$ after 10 years of observations}
\label{Fig:4}
\end{figure*}

\subsection{HESS J1837$-$069}
The H.E.S.S. Galactic Plane Survey (HGPS; \cite{Deil15}) in this region above 500 GeV shows a large and bright source, dubbed HESS J1837$-$069 \cite{Aharonian05,Aharonian06}. 
The source HESS J1837$-$069 is found to be associated with a spin-down pulsar PSR J1838$-$065 of characteristic age $t \approx 2.3\times 10^4$ year, surface dipole magnetic field $B_s \approx 1.9\times 10^{12}$ G and pulsation period $P_t \approx 70.5$ ms \cite{Gotthelf08}. Using the relation given in \cite{Bhadra06,Bednarek04}, we find that the initial period of the pulsar was about few millisecond only and therefore initial rotation energy of the pulsar was of the order of $10^{51}$ ergs which satisfies the required energy budget to interpret the observed radiations in hadronic interactions of so accelerated cosmic rays. A fraction of rotational energy released during slowing down of pulsar can be assumed to accelerate cosmic ray protons extracted from the surface of the neutron star \cite{Bhadra06,Bednarek04,Bartko08}. The accelerated primary cosmic rays may interact with the ambient medium and produce high energy gamma-rays and neutrinos. The primary cosmic-ray production spectrum has inferred by assuming released rotational energy of $E_{rot} = 3\times 10^{51}$ erg with a spectral index of $\alpha = -1.9$ using Eq.~\ref{Lcr}. Here, we consider an ambient matter density of the nebula of $n_H = 8.7$ cm$^{-3}$ to meet the observed gamma-ray spectrum. We notice that the efficiency of conversion of rotational energy to cosmic ray energy of $\xi = 10\%$ for primary cosmic ray protons can explain well the observed data.

The expected differential gamma-ray flux reaching the Earth from the PWN has presented in Fig.~\ref{Fig:4} along with the observations. We find that this hadronic model can also explain the observed SED from the source considering the maximum energy attainable by a cosmic ray proton in the PWN to be $E_{p, max} = 70$ TeV. As shown in Fig.~\ref{Fig:4}, future telescopes like e$-$ASTROGAM \cite{Angelis18}, CTA \cite{Ong17} and LHAASO \cite{Liu17}, which have higher sensitivity than the present generation gamma ray telescopes, should able to detect the EM SED of the source in a wide energy band and hence can help to identify the true physical model that leads to its EM emission. We have also considered the effect of background gamma-ray flux produced due to the interaction of galactic diffuse cosmic rays with the matter of the PWN. The mass of the gas content for the source is found out to be $\sim 2.8\times 10^{5}$ M$_{\odot}$ which is adopted to explain the observed gamma-ray spectrum.
Along with the EM SED matching by modeling, we have also reproduced the observed integral gamma-ray flux above $200$ GeV from the source as reported in \cite{Aharonian06} in order to precise determination of cosmic ray production spectral index $\alpha$ and also the maximum attainable energy $E_{max}$.

We have also determined the neutrino flux originated due to the decay of charged pions produced in the hadronic interaction of cosmic rays accelerated by HESS J1837$-$069 with the ambient matter (proton). The expected all flavor neutrino flux reaching at the Earth from the source is shown in the right-hand panel of Fig.~\ref{Fig:4}. We have estimated the expected muon-neutrino events in the IceCube detector from the PWN following \cite{Banik19} and it is found out to be $N_{\nu_{\mu}} \sim 1$ event in one year of observations above 1 TeV energies. We found that the estimated all flavor neutrino flux around 1 TeV energies is well within the reach of the sensitivity of IceCube-Gen2, a future extension of IceCube detector \cite{Aartsen19}.

\subsubsection{MAGIC J1835$-$069}
\begin{figure}
  \begin{center}
  \includegraphics[width = 0.48\textwidth,height = 0.45\textwidth,angle=0]{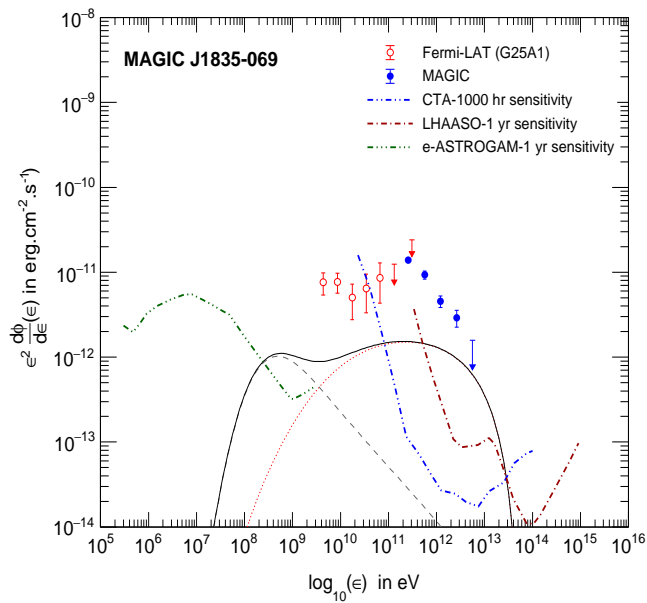}
\end{center}
  \caption{Same as in left-hand panel Fig.~\ref{Fig:4} but from MAGIC J1835$-$069 which is investigated to be a molecular cloud associated with HESS J1837$-$069.}
\label{Fig:5}
\end{figure}

The projected distance of the gamma-ray source MAGIC J1835$-$069 from the pulsar associated with HESS J1837$-$069 (for a distance of 6.6 kpc) is reported to be more than $\sim 65$ pc. As discussed above, it is unlikely that MAGIC J1835$-$069 is correlated with SNR G24.7+0.6. Here we have explored the possibility of molecular cloud origin of the observed gamma rays from MAGIC J1835$-$069, associated with the source HESS J1837$-$069. The energy density of runaway accelerated cosmic rays at the location of MAGIC J1835$-$069 coming from HESS J1837$-$069 can be obtained from Eq.~\ref{CR_den_cl} consistently where we have adopted the age of HESS J1837$-$069 to be $t = 2.3\times 10^4$ years \cite{Gotthelf08},  a diffusion constant of $D_0 = 2\times 10^{27}$ cm$^2$ s$^{-1}$ with $\delta = 0.33$ for cosmic rays. The distance of the molecular cloud from HESS J1837$-$069 is taken as $65$ pc following the MAGIC observation and the mass of the cloud is considered as $M_{cl} = 1.9\times 10^5$ M$_{\odot}$ (estimated for a distance of 6.6 kpc). The estimated differential gamma-ray flux reaching the Earth from MAGIC J1835$-$069  produced in hadronic interaction of these runaway cosmic rays with molecular cloud materials (protons) is shown in the Fig.~\ref{Fig:5} along with the observations. Our findings suggest that MAGIC J1835$-$069 is also unlikely to be associated with HESS J1837$-$069. 

\begin{figure*}
  \begin{center}
  \includegraphics[width = 1.0\textwidth,height = 0.45\textwidth,angle=0]{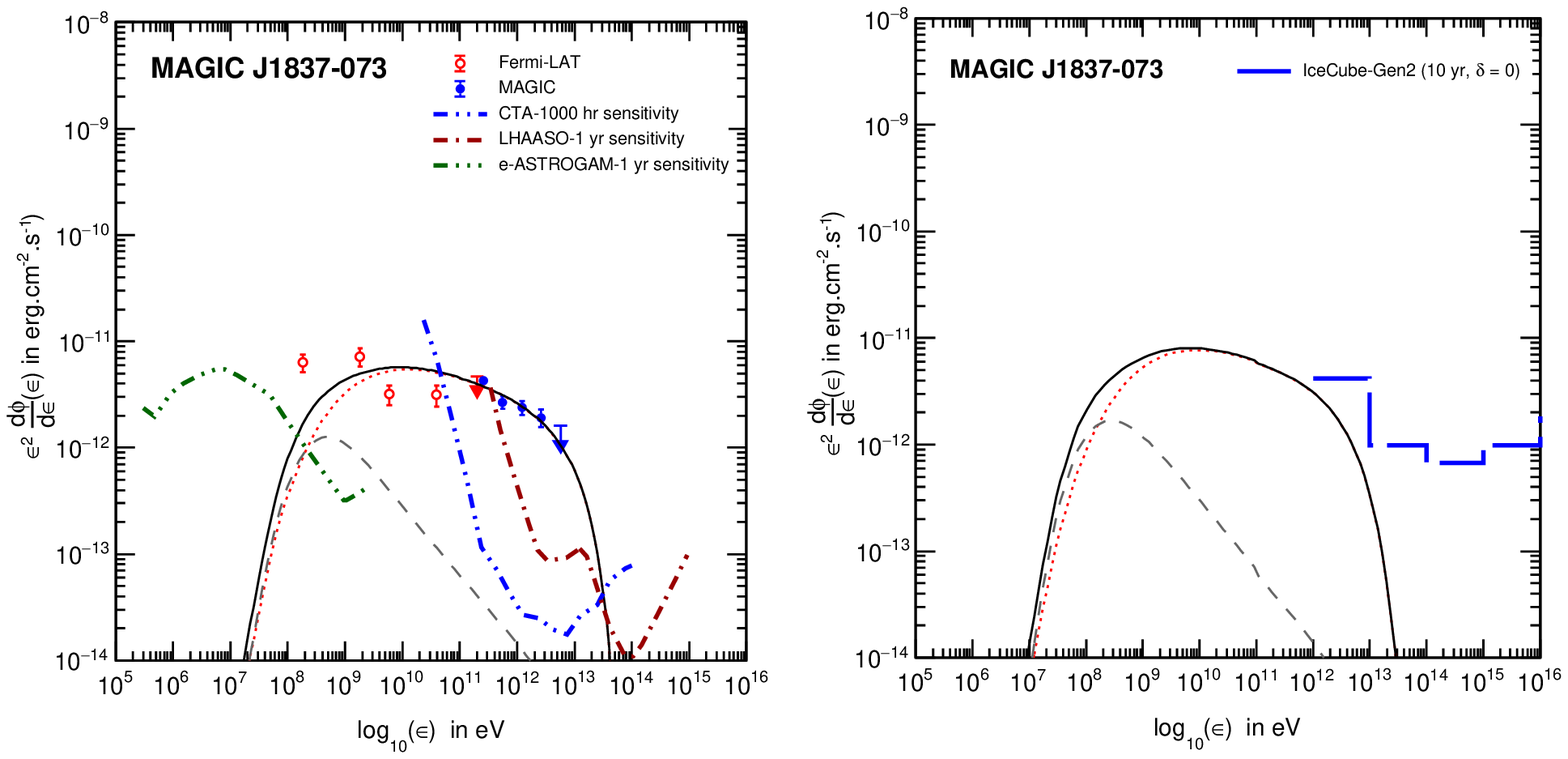}
\end{center}

  \caption{Same as in Fig.~\ref{Fig:4} but for MAGIC J1837$-$073 which is investigated to be a molecular cloud associated with HESS J1837$-$069.}
\label{Fig:6}
\end{figure*}

\subsubsection{MAGIC J1837$-$073}


The energy density of escaped accelerated cosmic rays at the position of MAGIC J1837$-$073 coming from HESS J1837$-$069 can be obtained from Eq.~\ref{CR_den_cl} where we consider a diffusion constant of $D_0 = 2\times 10^{27}$ cm$^2$ s$^{-1}$ with $\delta = 0.33$ for cosmic rays. The distance of the molecular cloud from HESS J1837$-$069 and mass of the cloud are adopted to be $38$ pc and $M_{cl} = 2.3\times 10^5$ M$_{\odot}$ respectively. The estimated differential gamma-ray flux reaching the Earth from MAGIC J1837$-$073 created in hadronic interaction of these runaway cosmic rays with molecular cloud materials (protons) is shown in the left panel of Fig.~\ref{Fig:6} along with the Fermi-LAT and MAGIC observations. Our findings suggest that MAGIC J1835$-$073 is likely to be associated with HESS J1837$-$069.

We have also estimated the neutrino flux originated due to the decay of charged pions created in the hadronic interaction of cosmic rays accelerated by HESS J1837$-$069 with the molecular cloud. The expected all flavor neutrino flux reaching at the Earth from the source is shown in the right-hand panel of Fig.~\ref{Fig:6}. We have estimated the corresponding expected muon-neutrino events in the IceCube detector from the SNR following \cite{Banik19} and it is found out to be $N_{\nu_{\mu}} = 0.21$ events in 1 year of observations above 1 TeV energies. We found that the estimated all flavor neutrino flux around 1 TeV energies may be detected by IceCube-Gen2 \cite{Aartsen19}.

\section{Conclusion}

On the basis of our present analysis we conclude the followings:

i) The Fermi-LAT has detected the SNR G24.7+0.6 in the GeV energy range but the  MAGIC telescope did not detect any TeV gamma rays from the source, instead they found a new TeV source, the MAGIC J1835$-$069, in the neighborhood of the SNR G24.7+0.6 resolving it from SNR G24.7+0.6 at a level of $13\sigma$. The observed gamma ray emission from the SNR G24.7+0.6 can be interpreted in terms of hadronic interaction of the SNR accelerated cosmic rays with the ambient matter but the non-detection of TeV gamma rays from the source restricts the maximum energy attainable by a cosmic ray particle in the source to just 1.3 TeV. Consequently, there will be no TeV neutrinos from the source. 

ii) The present analysis suggests that the GeV-TeV emission of HESS J1837$-$069 can well be described by hadronic interaction of the PWN accelerated cosmic rays with the ambient matter. The maximum energy attainable by a cosmic ray proton in the source turns out to be 70 TeV which can be verified in the future by the upcoming CTA and LHAASO detectors. The estimated all flavor neutrino flux above 1 TeV energy is found to be well within the reach of the sensitivity of IceCube-Gen2, a future extension of IceCube detector. 

iii) The newly detected source MAGIC J1835$-$069 is unlikely to be a cosmic ray illuminated molecular cloud. The present analysis shows that the runway cosmic rays either from the SNR G24.7+0.6 or the PWN HESS J1837$-$069 fall short to reproduce the observed TeV emission from MAGIC J1835$-$069 by interacting with molecular cloud at the location of MAGIC J1835$-$069. Instead of considering the SNR G24.7+0.6 and MAGIC J1835$-$069 as a resolved separate sources recently Sun et al. \cite{Sun20} have considered the gamma ray emission from the two sources together (over an extended region) and described the observed GeV-TeV gamma rays from the region containing SNR G24.7+0.6 and MAGIC J1835$-$069 by hadronic interaction of cosmic rays with ambient matter (cloud). Since the MAGIC telescope resolved MAGIC J1835$-$069 from the SNR G24.7+0.6 with high significance we have here treated the stated sources separately. It may happen that MAGIC J1835$-$069 is the major part of the ejecta of supernova explosion that leads to G24.7+0.6. 

iv)  The newly detected TeV gamma ray source MAGIC J1837$-$073 appears to be associated with the PWN HESS J1837$-$069. The gamma ray emission from MAGIC J1837$-$073 can well be interpreted by hadronic interactions of runway cosmic rays from HESS J1837$-$069 with the molecular cloud at the location of MAGIC J1837$-$073. The flux of TeV neutrinos from the source is, however, too small to be detected by the present or any upcoming neutrino telescope.

\section*{Acknowledgements}
The authors would like to thank an anonymous reviewer for insightful comments and useful suggestions that helped us to improve the manuscript. AB acknowledges the financial support from SERB (DST), Govt. of India vide approval number CRG/2019/004944.


\end{document}